\documentclass[fleqn,12pt,twoside]{article}
\usepackage{espcrc1}

\renewcommand{\topmargin}{-18mm}

\usepackage{graphicx}
\usepackage[figuresright]{rotating}

\newcommand{\AmS}{{\protect\the\textfont2
  A\kern-.1667em\lower.5ex\hbox{M}\kern-.125emS}}
\newcommand{\case}[2]{\mbox{$\textstyle{#1\over#2}$}}

\title{Quantum Monte Carlo Calculations of Light Nuclei}

\author{Steven C. Pieper 
\address[]{Physics Division, Argonne National Laboratory\\
        Argonne, Illinois 60439, USA}
\thanks{Work supported by the United States Department of Energy, 
Office of Nuclear Physics, under contract No.~W-31-109-ENG-38.
The many-body calculations were performed on the parallel computers of the
Laboratory Computing Resource Center, Argonne National Laboratory,
the National Energy Research Scientific Computing Center,
and Los Alamos National Laboratory.
This work has been done with J.~Carlson, D.~Kurath, V.~R.~Pandharipande,
K.~Varga, and R.~B.~Wiringa.}
}
       
\begin{document}

\maketitle

\begin{abstract}

Variational Monte Carlo and Green's function Monte Carlo are
powerful tools for calculations of properties of light nuclei 
using realistic two-nucleon ($N\!N$) and
three-nucleon ($N\!N\!N$) potentials.
Recently the GFMC method has been extended to multiple states
with the same quantum numbers.
The combination of the Argonne $v_{18}$ two-nucleon and Illinois-2
three-nucleon potentials gives a good prediction of many energies
of nuclei up to $^{12}$C.  
A number of other recent results are presented: comparison
of binding energies with those obtained by the no-core shell model;
the incompatibility of modern nuclear Hamiltonians with a bound
tetra-neutron; difficulties in computing RMS radii of very weakly
bound nuclei, such as $^6$He; center-of-mass effects on spectroscopic
factors; and the possible use of an artificial external well in 
calculations of neutron-rich isotopes.

\end{abstract}

\section{INTRODUCTION}

In the last decade, the Green's function Monte Carlo (GFMC) method has
been developed into a powerful tool for calculations of light nuclei (so
far up to $A$~=~12) using realistic two-nucleon ($N\!N$) and
three-nucleon ($N\!N\!N$) potentials.  GFMC starts with a trial wave
function that is obtained via variational Monte Carlo (VMC) and projects
out excited state contamination to, in principle, obtain the true
ground-state wave function for the given Hamiltonian.  In practice the
method obtains ground and low-lying excited state energies with an
accuracy of 1--2\%.  A review of the nuclear VMC and GFMC methods up to
$A$~=~8 may be found in Ref.~\cite{PW01}; $A$~=~9,10 results are in
Ref.~\cite{PVW02}.  By using the Argonne $v_{18}$ $N\!N$ potential
(AV18) and including two- and three-pion exchange $N\!N\!N$ potentials,
a series of model Hamiltonians (the Illinois models) were
constructed~\cite{PPWC01} that give a good reproduction of energies for
$A$~=~3 to 12. 
Recently GFMC has been extended to the calculation of multiple
excited states with the same quantum numbers~\cite{PWC04}.

This paper reviews some of these results and compares GFMC energies
to No-core Shell Model (NCSM) results~\cite{nocore} for several cases.  
A recent study showing that
an experimental claim of a bound tetraneutron is very
unlikely to be valid~\cite{P03} is also reprised.  Finally some
on-going work involving
spectroscopic factors, neutron-rich oxygen isotopes, 
and difficulties in GFMC computation of RMS radii
is presented.

\section{HAMILTONIANS}

Our Hamiltonian includes a nonrelativistic one-body kinetic energy, the
Argonne $v_{18}$ (AV18) two-nucleon potential \cite{WSS95} and various
three-nucleon potentials,
\begin{equation}
H = \sum_{i} (-\frac{\hbar^2}{2m} \nabla^{2}_{i}) + \sum_{i<j} v_{ij}
 + \sum_{i<j<k} V_{ijk} \ .
\end{equation}
The difference between proton and neutron masses is included in our
calculations, but ignored above.  The Argonne $v_{18}$ potential is one
of a number of accurate $NN$ potential models developed
since 1990.  It can be written as a sum of electromagnetic and
one-pion-exchange terms and a shorter-range phenomenological part.
The electromagnetic terms include one- and two-photon-exchange Coulomb
interactions, vacuum polarization, Darwin-Foldy, and magnetic moment
terms, with appropriate proton and neutron form factors.

The one-pion-exchange part contains the normal Yukawa and tensor
functions with a short-range cutoff.  This and the remaining
phenomenological part of the potential can be written as a sum of
eighteen operators, which is where the name $v_{18}$ comes from.
The first fourteen are charge-independent, and include spin-spin, tensor,
$L \cdot S$, and quadratic-$L$ terms, each with a dependence on isospin.
The last four operators break charge independence.  The radial forms 
associated with each operator are
determined by fitting $N\!N$ scattering data.
The potential was fit directly to the Nijmegen $N\!N$ scattering data
base~\cite{Nijmegen}, which contains 1787 $pp$ and 2514 $np$
data in the range $0-350$ MeV, with a $\chi^2$ per datum of 1.09.  It
was also fit to the $nn$ scattering length measured in
$d(\pi^-,\gamma)nn$ experiments and the deuteron binding energy.

\begin{figure}[h] 
\begin{minipage}[t]{80mm}
\includegraphics[height=1.5in]{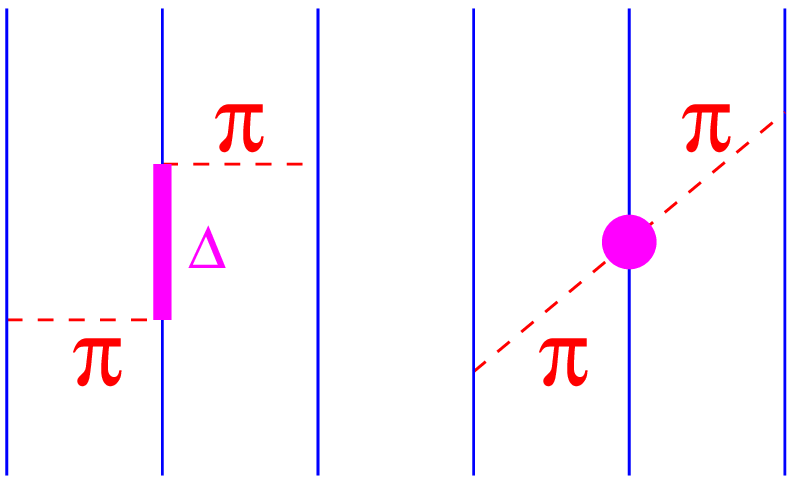}
\end{minipage}
\hspace{\fill}
\begin{minipage}[t]{75mm}
\includegraphics[height=1.5in]{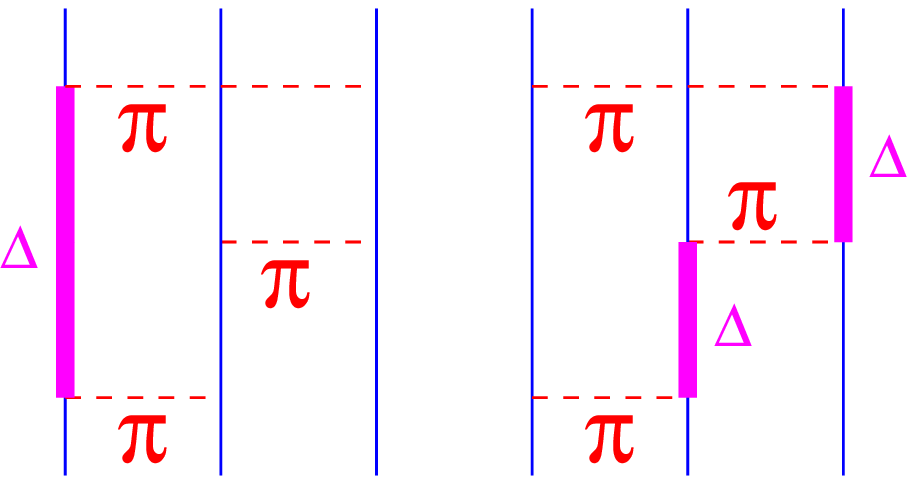}
\end{minipage}
\caption{Terms in the Illinois three-nucleon potentials.}
\label{fig:vijk}
\end{figure}

The Urbana series of three-nucleon
potentials were developed to compute properties of $A=3,4$ nuclei and
nuclear matter; the current (ninth) model is designated UIX~\cite{PPCW95}.  
These potentials are written as sums of two-pion-exchange with
intermediate excitation of an isobar (left panel of Fig.~\ref{fig:vijk}) 
and shorter-range phenomenological terms.  
The two-pion-exchange term is that of the original
Fujita-Miyazawa model~\cite{FM57} and contains both spin (tensor) and
isospin dependence.  The shorter-range phenomenological term
is purely central and repulsive.  
Our recent work has shown the need for additional binding for
$p$-shell nuclei and for further increased binding as $|N-Z|$ increases.
This led to the development of the Illinois models
\cite{PPWC01} which, in addition to the Urbana terms, contain the
two-pion $s$-wave scattering term (second panel of Fig.~\ref{fig:vijk})
and three-pion exchange ring terms (last two panels of Fig.~\ref{fig:vijk}).
The latter can involve the excitation of one or two sequential isobars,
so that each energy denominator contains only one $\Delta$ mass.

In light nuclei we find that the three-nucleon potential contributes
only 2-9\% (increasing with $A$) of the total potential energy.
However, due to the large cancellation of potential and kinetic energy,
this amounts to 15-50\% (increasing mostly with $N-Z$) of the binding
energy.  We expect a similar ratio for the four-body potential, which
implies that it contributes only a few percent of the binding energy;
such contributions are close to our computational accuracy and
would be absorbed in fit of the $N\!N\!N$ potential parameters.

\section{QUANTUM MONTE CARLO METHODS}

\subsection{Variational Monte Carlo
}
Variational Monte Carlo finds an upper bound, $E_T$, to an eigenenergy
of $H$ by evaluating the expectation value of $H$ in a trial wave
function, $\Psi_T$.  The parameters in $\Psi_T$ are varied to minimize
$E_T$, and the lowest value is taken as the approximate energy.  Over
the years, we have developed rather sophisticated $\Psi_T$ for light
nuclei~\cite{PPCPW97,Wir00}.  
A good variational trial function has the form
\begin{equation}
     |\Psi_T\rangle = \left[1 + \sum_{i<j<k} U^{TNI}_{ijk} \right]
                      \left[ {\mathcal S}\prod_{i<j}(1+U_{ij}) \right]
                      |\Psi_J\rangle \ .
\label{eq:psit}
\end{equation}
The $U_{ij}$ and $U^{TNI}_{ijk}$ are non-commuting two- and three-nucleon
correlation operators (the most important being
the tensor-isospin correlation corresponding to the pion-exchange
potential); ${\mathcal S}$ indicates a symmetric sum over
all possible orderings; and $\Psi_J$ is a fully antisymmetric Jastrow
wave function which
determines the quantum numbers of the state being computed.

The Jastrow wave function, $\Psi_J$, for $p$-shell nuclei
starts with a sum over independent-particle terms, $\Phi_A$, that have 4
nucleons in an $\alpha$-like core and $(A-4)$ nucleons in $p$-shell orbitals.
These orbitals are coupled in a $LS[n]$ basis
to obtain the desired $JM$ value of a given state, where $n$ specifies
the spatial symmetry $[n]$ of the angular
momentum coupling of the $p$-shell nucleons.
Different possible $LS[n]$ combinations lead to multiple components in the
Jastrow wave function.
This independent-particle basis is acted on by products of central pair
and triplet correlation functions:
\begin{eqnarray}
  |\Psi_{J}\rangle &=& {\mathcal A} \left\{
     \Big[\prod_{i<j<k} f^{c}_{ijk}\Big] \Big[\prod_{i<j \leq 4} f_{ss}(r_{ij})\Big]
     \sum_{LS[n]} \Big( \beta_{LS[n]}
     \Big[\prod_{k \leq 4<l \leq A} f^{LS[n]}_{sp}(r_{kl})\Big] \right.   \nonumber\\
  &\times& \left. \Big[\prod_{4<l<m \leq A} f^{LS[n]}_{pp}(r_{lm})\Big]
    |\Phi_{A}(LS[n]JMTT_{3})_{1234:5\ldots A}\rangle \Big) \right\} \ .
\label{eq:psip}
\end{eqnarray}
The operator ${\mathcal A}$ indicates an antisymmetric sum over all possible
partitions of the $A$ particles into 4 $s$-shell and $(A-4)$ $p$-shell ones.
The pair correlation for both particles within the $s$-shell, $f_{ss}$,
is similar to that in an $\alpha$-particle.
The pair correlations for both particles in the $p$-shell, $f^{LS[n]}_{pp}$,
and for mixed pairs, $f^{LS[n]}_{sp}$, are similar to $f_{ss}$ at short
distance, but their long-range structure is adjusted to give appropriate
clustering behavior, and they may vary with $LS[n]$.

The single-particle wave functions $\Phi_A$ are given by:
\begin{eqnarray}
 &&  |\Phi_{A}(LS[n]JMTT_{3})_{1234:5\ldots A}\rangle =
     |\Phi_{4}(0 0 0 0)_{1234} \Big[\prod_{4 < l\leq A}
     \phi^{LS[n]}_{p}(R_{\alpha l})\Big] \\
 && \times \left\{ \Big[ \prod_{4 < l\leq A} Y_{1m_l}(\Omega_{\alpha l}) \Big]_{LM_L[n]}
     \Big[ \prod_{4 < l\leq A} \chi_{l}(\frac{1}{2}m_s) \Big]_{SM_S}
     \right\}_{JM}
     \Big[ \prod_{4 < l\leq A} \nu_{l}(\frac{1}{2}t_3) \Big]_{TT_3}\rangle
     \nonumber \ .
\label{Phi_A}
\end{eqnarray}
The $\phi^{LS[n]}_{p}(R_{\alpha l})$ are $p$-wave solutions of a particle
in an effective $\alpha-N$ potential that has Woods-Saxon and Coulomb parts.
They are functions of the distance between the center of mass
of the $\alpha$ core and nucleon $l$, and may vary with $LS[n]$.
The depth, width, and surface thickness of the single-particle potential 
are additional variational parameters of the trial function.
The overall wave function is translationally invariant, so there is no
spurious center of mass motion.

The $\beta_{LS[n]}$ mixing parameters of Eq.~(\ref{eq:psip}) are determined
by a diagonalization procedure, in which matrix elements
\begin{eqnarray}
       E_{T,ij} &=& \langle \Psi_T(\beta_i) | H | \Psi_T(\beta_j) \rangle \ , 
\label{eq:geneigen-e}\\
       N_{T,ij} &=& \langle \Psi_T(\beta_i) | \Psi_T(\beta_j) \rangle \ ,
\label{eq:geneigen-n}
\end{eqnarray}
are evaluated using trial functions $\Psi_T(\beta_i) \equiv \Psi_T(\beta_i=1,
\beta_{j\ne i}=0)$.
Although the $\Phi_A(LS[n]JMTT_3)$ are orthogonal due to spatial symmetry,
the pair and triplet correlations in $\Psi_T$ make the different $LS[n]$
components nonorthogonal, so a generalized eigenvalue routine is necessary
to carry out the diagonalization.

\subsection{Green's Function Monte Carlo}

GFMC projects out the lowest-energy ground state from the VMC $\Psi_T$ by using 
\begin{eqnarray}
\Psi(\tau) &=& e^{-({H}-E_{0})\tau} \Psi_{T} ~;   \\
\Psi_0 &=& \lim_{\tau \rightarrow \infty} \Psi(\tau) .
\end{eqnarray}
If sufficiently large $\tau$ is reached,
the eigenvalue $E_{0}$ is calculated exactly while other expectation values
are generally calculated neglecting terms of order $|\Psi_{0}-\Psi_{T}|^{2}$ 
and higher.  
In contrast, the error in the variational energy, $E_{T}$, is of order 
$|\Psi_{0}-\Psi_{T}|^{2}$, and other expectation values calculated with 
$\Psi_{T}$ have errors of order $|\Psi_{0}-\Psi_{T}|$.
Here I present a simplified overview of nuclear GFMC;
a rather complete discussion may be found in \cite{PPCPW97,Wir00}.

Introducing a small time step, $\triangle\tau$, $\tau=n\triangle\tau$, gives
(typically $\triangle\tau = 0.5$ GeV$^{-1}$)
\begin{eqnarray}
\Psi(\tau) = \left[e^{-({H}-E_{0})\triangle\tau}\right]^{n} \Psi_{T}
= G^n \Psi_{T} \ .
\end{eqnarray}
where $G$ is the short-time Green's function.
The $\Psi (\tau)$ is represented by a vector function of $\bf R$, and the
Green's function, $G_{\alpha\beta}({\bf R}^{\prime},{\bf R})$ is a matrix
function of $\bf R^{\prime}$ and ${\bf R}$ 
in spin-isospin space (labeled by the subscripts $\alpha,\beta$), defined as
\begin{eqnarray}
G_{\alpha\beta}({\bf R}^{\prime},{\bf R})= \langle {\bf 
R}^{\prime},\alpha|e^{-({H}-E_{0})\triangle\tau}|{\bf R},\beta\rangle.
\label{eq:gfunction}
\end{eqnarray}
Omitting
spin-isospin indices for brevity, $\Psi({\bf R}_{n},\tau)$ is given by
\begin{eqnarray}
\Psi({\bf R}_{n},\tau) = \int G({\bf R}_{n},{\bf R}_{n-1})\cdots G({\bf 
R}_{1},{\bf R}_{0})\Psi_{T}({\bf R}_{0})d{\bf R}_{n-1}\cdots d{\bf R}_{1}d{\bf 
R}_{0} ,
\label{eq:gfmcpsi}
\end{eqnarray}
and the integration is done by Monte Carlo.
We approximate the short-time propagator as a symmetrized product of
exact two-body propagators and include the $V_{ijk}$ to first-order.
In recent benchmark calculations~\cite{KNG+01} of $^4$He
using an eight-operator NN potential, the GFMC energy had a statistical
error of only 20 keV and agreed with the other best results to this
accuracy ($<$ 0.1\%).

For more than four nucleons, GFMC calculations suffer significantly from
the well-known Fermion sign problem.  This results in exponential growth of the statistical errors as one
propagates to larger $\tau$, or as $A$ is increased.
For $A \geq 8$ the resulting limit on  $\tau$ is too small to
allow convergence of the energy.
In the last few years we have developed and extensively tested a
con\-strain\-ed-path algorithm for nuclear GFMC~\cite{Wir00}
to solve this problem.  In this
method configurations with small or negative
$\Psi(\tau)^\dagger\!\cdot\!\Psi_T$ are discarded such that the average over
all discarded configurations of $\Psi(\tau)^\dagger\cdot\Psi_T$ is 0.
This means that, if $\Psi_T$ were the true eigenstate, the discarded
configurations would contribute nothing but noise to $\langle H
\rangle$.  This constrained propagation completely controls the growth
of the statistical errors and in most cases produces a result that is
statistically the same as unconstrained propagation (the accuracy of the
comparison may be limited by the statistical errors in the unconstrained
result).  However we have demonstrated some cases for which constrained
propagation leads to a wrong result, and in fact for which the
approximate $\langle H \rangle$ is not even an upper bound to the
correct eigenvalue.  In all cases the correct result can be obtained by
making a few (10 to 20) unconstrained steps before evaluating the energy.
Our calculations for $A \geq 5$ are now all made using constrained-path
propagation with 10 to 20 unconstrained steps.

The number of spin-isospin components in $\Psi_T$ grows rapidly with the
number of nucleons.  Thus a calculation of a state in $^8$Be involves
about 30 times more floating-point operations than one for $^6$Li, and
$^{10}$Be requires 50 times more than $^8$Be.  Calculations of the sort
being described here are currently feasible up to only $A = 12$; 
for $A$=10, these
require $\sim$8,000 processor hours on the NERSC IBM SP (Seaborg)
running at 390 MFLOPS/processor ($10^{16}$ operations)
and $\sim$150,000 processor hours for $^{12}$C on the Los Alamos
qsc computer running at 360 MFLOPS/processor ($2\times10^{17}$ operations).

\begin{figure}[htb!]
\centering
\includegraphics[width=4.0in,angle=270]{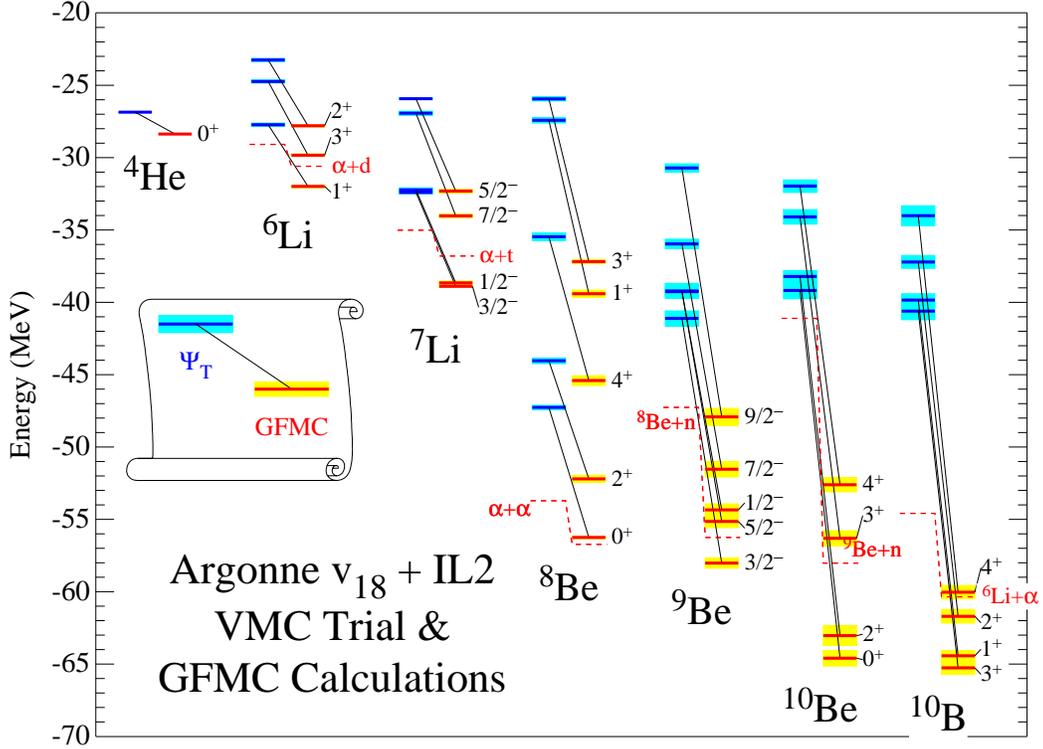}
\caption{Comparison of $\Psi_T$ and GFMC energies for several nuclei}
\label{fig:tg}
\end{figure}

Figure \ref{fig:tg} compares the VMC and GFMC energies of several
nuclei for the AV18+IL2 Hamiltonian.  We see that the variational
wave functions for the $s$-shell nuclei are quite accurate--the
GFMC improves our best VMC energy of $^4$He by only 1.5~MeV or 5\%.
(These $\Psi_T$ do not contain explicit correlations for
the new terms in the Illinois $N\!N\!N$ potential; the corresponding
error for the AV18+UIX Hamiltonian is only 2\%.)  However,
the $p$-shell variational wave functions are much less accurate; compared
to the GFMC energies, they result in underbindings 
of 4.3 MeV (13\%) for $^6$Li to 25 MeV (38\%) for $^{10}$B.
In fact the $\Psi_T$ fail to give particle-stable energies for
any of the $A\geq 6$ nuclei and give maximum binding energies
for $^8$Be.
On the other hand, the excitation spectra
from VMC and GFMC calculations are generally quite similar,
except when there is a change in the dominant symmetry.

\subsection{GFMC Evaluation of Excited States}

It is possible to treat at least a few excited states with the
same quantum numbers using VMC and GFMC methods~\cite{PWC04}.
The VMC calculations have been described above, and essentially
involve solving a generalized eigenvalue problem, Eqs.~(\ref{eq:geneigen-e})
and (\ref{eq:geneigen-n}).
The same basic method can be applied in GFMC calculations, though
the implementation is slightly more involved.
In this section, $\Psi_{T,i}$ represents the trial wave function for
the $i^{th}$ state of specified $(J^\pi,T)$ and $\Psi_i(\tau)$ is
the GFMC wave function propagated from it.  By construction
$\langle \Psi_{T,i} | \Psi_{T,j} \rangle = 0$ for $i \neq j$.
We would like to calculate the Hamiltonian and
normalization matrix elements as a function of $\tau$:
\begin{eqnarray}
H_{ij}(\tau) & = & \frac {\langle \Psi_i(\tau/2)| \ H \ | \Psi_j(\tau/2) \rangle}
 {|\Psi_i(\tau/2)| |\Psi_j(\tau/2)|} ,  \\
N_{ij} (\tau) & = & \frac {\langle \Psi_i(\tau/2) | \Psi_j(\tau/2) \rangle}
 {|\Psi_i(\tau/2)| |\Psi_j(\tau/2)|} ,
\end{eqnarray}
where $| \Psi_i | = | \langle \Psi_i | \Psi_i \rangle |^{1/2}$.
Solving the generalized eigenvalue problem with
these Hamiltonian and normalization matrix elements would yield
improved upper bounds for the ground and low-lying excited states
of the system.  In the limit $\tau \rightarrow \infty $ the
solutions would be exact.

In GFMC we can compute mixed expectation values such as
\begin{equation}
 \tilde{O}_{ij}(\tau) = \frac {\langle \Psi_i(\tau)| \ O \ | \Psi_{T,j} \rangle}
 {\langle \Psi_i(\tau)| \Psi_{T,i} \rangle} ,
\end{equation}
where the denominator involves just state $i$.  
Since the propagator commutes with the Hamiltonian, the desired matrix
elements can be computed as:
\begin{eqnarray}
H_{ij} (\tau) & = & [ \tilde{H}_{ij} \tilde{H}_{ji} ]^{\case{1}{2}} , \\
N_{ij} (\tau) & = & [ \tilde{N}_{ij} \tilde{N}_{ji} ]^{\case{1}{2}} ,
\end{eqnarray}
where we use expectation values computed from separately propagated $\Psi_i(\tau)$
and $\Psi_j(\tau)$.  For $i=j$ these equations reduce to the standard
GFMC calculation described above.

\begin{figure}[b!]
\centering
\includegraphics[width=4.0in,angle=270]{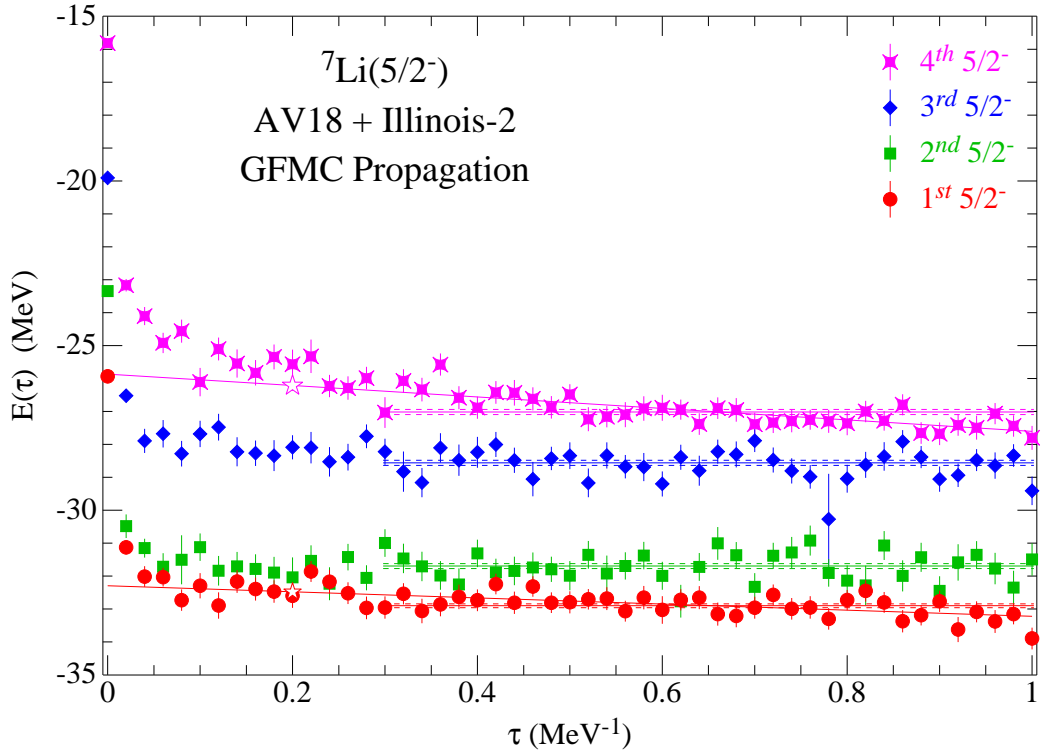}
\caption{GFMC energies of four $\frac{5}{2}^-$ states in $^7$Li versus 
imaginary time, $\tau$.}
\label{fig:e-vs-tau}
\end{figure}

As an example, Fig. \ref{fig:e-vs-tau} shows
the computation of the energies of four $\case{5}{2}^-$ states in $^7$Li.
The lowest state has mainly [43] symmetry and can easily decay to the
$\alpha$+t channel; it has a large experimental width (918 keV) and its
computed energy is slowly decreasing to the energy of the separated clusters.
The remaining states are mainly of [421] symmetry and so are principally
connected to the $^6$Li+n channel.  The second $\case{5}{2}^-$ state is
experimentally just above the $^6$Li+n threshold and has a small
width (80 keV); its computed energy becomes constant with increasing $\tau$.
The last two $\case{5}{2}^-$ states are not experimentally known,
but the very slow decrease with $\tau$ of the energy of the third state
suggests that this state may also be narrow.  
The off-diagonal overlaps $N_{i1}(\tau)$ are
small and do not show signs of steadily increasing with increasing $\tau$.
The solutions of generalized eigenvalue problems using the $E_{ij}(\tau)$
and $N_{ij}(\tau)$ are not significantly different from the 
$E_{ii}(\tau)$ shown in the figure.
These results show that the (constrained) GFMC propagation largely
retains the orthogonality of the starting $\Psi_{T,i}$.
Contrary to what might have been expected, the propagation of the
higher states does not quickly collapse to the lowest state.

\section{ENERGIES OF NUCLEAR STATES}

\begin{figure}[tb!]
\centering
\includegraphics[height=6.30in,angle=270]{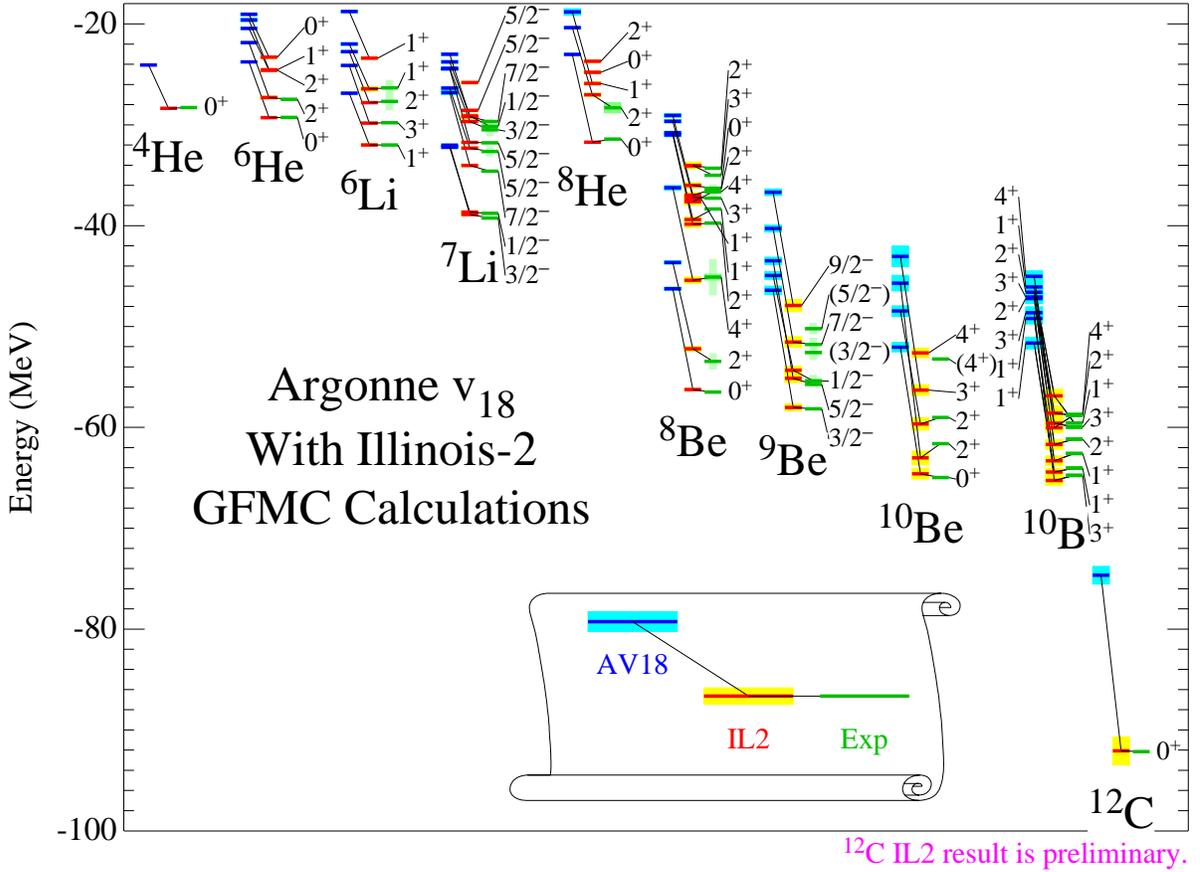}
\caption{Energies of nuclear states computed with just the AV18 $N\!N$
potential, and with the addition of the IL2 $N\!N\!N$ potential, 
compared to experiment.}
\label{fig:energies}
\end{figure}

\begin{figure}[tb!]
\centering
\includegraphics[height=6.30in,angle=270]{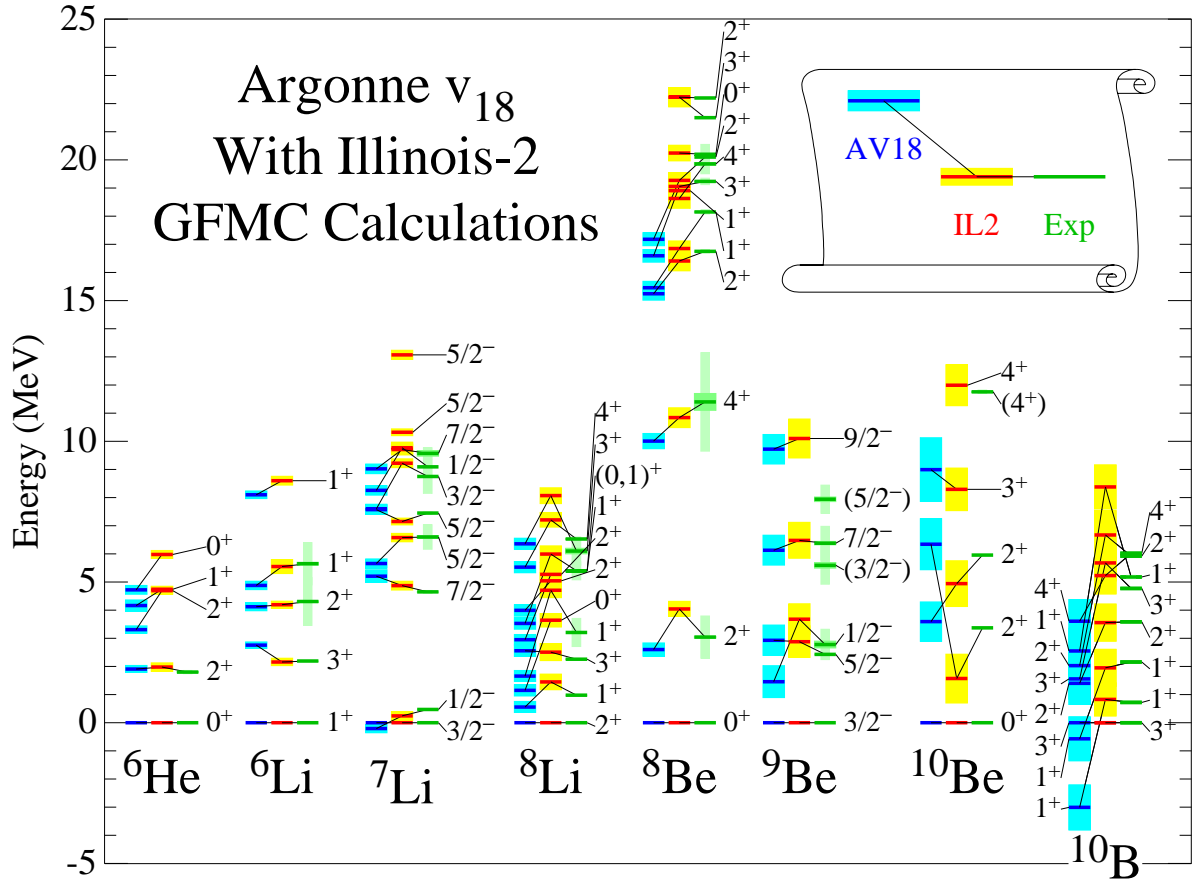}
\caption{Excitation energies of nuclear states.}
\label{fig:estar}
\end{figure}

Figure \ref{fig:energies} compares energies computed with
the AV18 (no $V_{ijk}$) and AV18+IL2 Hamiltonians to experimental
values.  The AV18+IL2 result shown for $^{12}$C was made using
a simplified $\Psi_T$ and an approximate treatment of $V_{ijk}$
in the GFMC propagation; for these reasons it is marked preliminary.
We see that using just a $N\!N$ potential underbinds
$^4$He by 4 MeV; this underbinding increases to 18 MeV for $^{12}$C.
The parameters of the Illinois-2 $N\!N\!N$ potential were 
adjusted to reproduce the energies of 17 narrow states for
$3 \leq A \leq 8$~\cite{PPWC01}.  As can be seen the potential provides an
excellent overall reproduction of the energies of many states
up to the ground state of $^{12}$C; the RMS error in reproducing
the experimental energies of the 39 states with width $< 0.2$~MeV
is 0.7~MeV.

Figure \ref{fig:estar} 
shows excitation spectra of nuclei.  One again sees a generally
good reproduction using AV18+IL2 of the experimental values.
The $N\!N\!N$ potential increases the predicted
splittings of several spin-orbit partners, such as the 
$3^+$,$2^+$,$1^+$ triplet in $^6$Li
and the first $\case{7}{2}^-$ and $\case{5}{2}^-$ levels in $^7$Li
which significantly improves the agreement with experiment.
A variational study of $^{15}$N also showed that a large part of
the spin-orbit splitting in that nucleus is a result of the
$N\!N\!N$ potential~\cite{PP93}.
However the most dramatic effect of the $N\!N\!N$ potential on levels 
appears in $^{10}$B.  The AV18 Hamiltonian results in two $1^+$ levels
being below the $3^+$ state which is the experimental ground state;
the AV18+IL2 Hamiltonian gives the correct order.
A similar inversion of levels occurs for the $\case{5}{2}^-$ and 
$\case{1}{2}^-$ states in $^9$Be.  
These inversions are also due to increased spin-orbit strength
from the $N\!N\!N$ potential; in 1956 Kurath showed that the position
of the $^{10}$B(3$^+$) state depends sensitively on spin-orbit 
strength~\cite{kurath56}.
Another
level ordering that is correct only with the $N\!N\!N$ potential 
is shown in $^{10}$Be.  Here, as can be inferred from the experimental
$BE(2)$'s to the ground state, the first $2^+$ level should have
a large negative quadrupole moment and the second should have
a positive quadrupole moment.  This is the case for the full
AV18+IL2 Hamiltonian but without IL2 the energies of these 
states are reversed.

\section{OTHER RESULTS}

\subsection{Comparison of GFMC and No-core Shell Model Calculations}

\begin{figure}[b!]
\centering
\includegraphics[height=6.30in,angle=270]{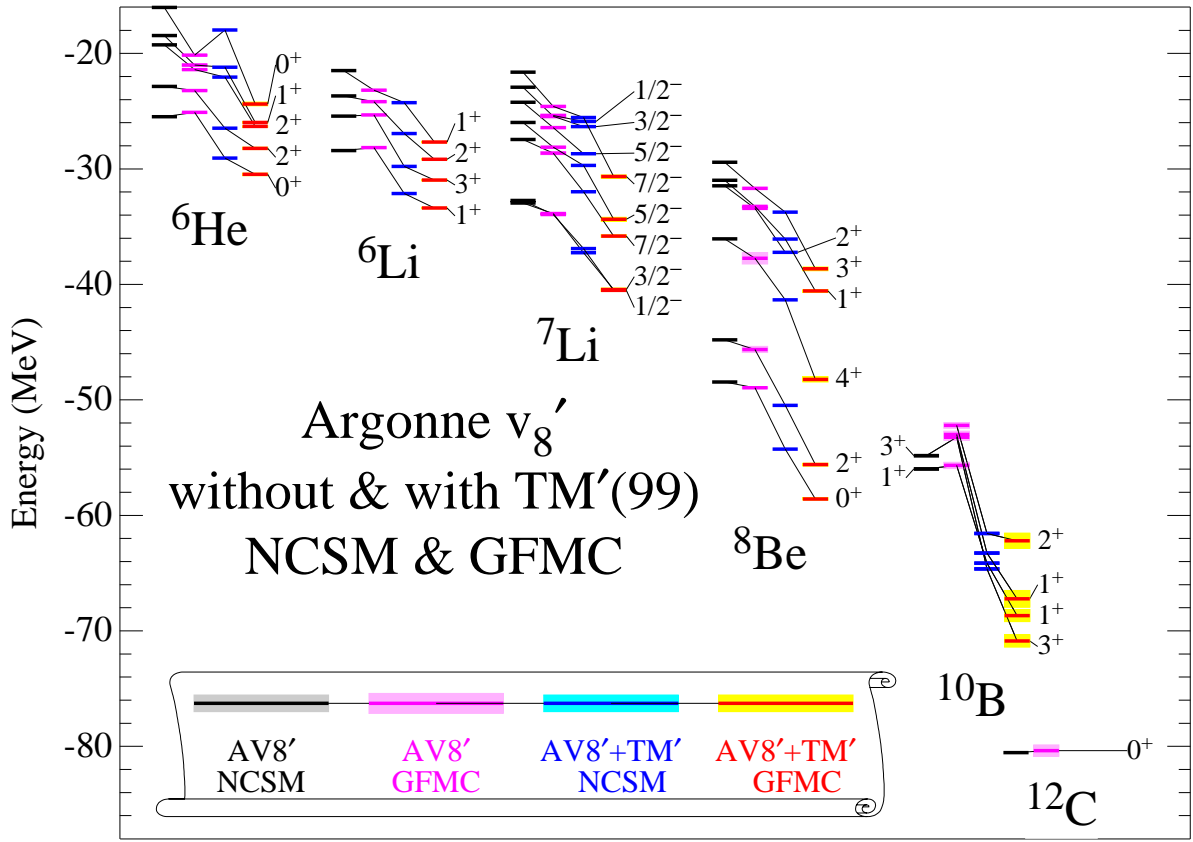}
\caption{Comparison of NCSM and GFMC energies for the AV8$^\prime$ and
AV8$^\prime$+TM$^\prime$ Hamiltonians.}
\label{fig:ncsm}
\end{figure}

The no-core shell model (NCSM) is an alternative many-body method that
uses realistic $N\!N$ and $N\!N\!N$ potentials for light nuclei.  It
uses an expansion in a large harmonic-oscillator basis for all $A$
nucleons.  Effective two- and three-body interactions are constructed
from the bare $v_{ij}$; bare $V_{ijk}$ can also be included in the
three-body effective interactions.  The construction of the effective
interactions also gives a unique prescription for constructing effective
operators from bare operators.  The size of the harmonic-oscillator
basis for a specified order of expansion grows rapidly with $A$.  At
present 16$\hbar\Omega$ can be used for $A=6$ with just two-body
effective interactions; this results in well-converged calculations.
However only 8$\hbar\Omega$ can be used for $A=12,16$ and two-body
effective interactions and these calculations are not fully converged.
Calculations with three-body effective interactions are limited to
6$\hbar\Omega$ and are generally not converged.  
For a complete
description of this method see~\cite{nocore} and references therein.

The NCSM is much faster than GFMC but because of these limits on the
basis size appears to be less accurate than GFMC, especially when a
$N\!N\!N$ potential is being used.  This is shown in Fig.~\ref{fig:ncsm}
in which NCSM and GFMC calculations of several nuclei are compared.  The
left-hand two bars for each state show, respectively, NCSM and GFMC
calculations for the AV8$^\prime$ potential~\cite{PPCPW97} with no 
$N\!N\!N$ potential;
there is good agreement between both methods.  The right-hand two
bars shows results with the TM$^\prime$(99) $N\!N\!N$ potential~\cite{tmp}
added to AV8$^\prime$.  In this case the NCSM results are significantly
above those from GFMC; that is the NCSM appears to not get the full
additional attraction provided by TM$^\prime$(99).  This is probably
because of the limitation to only 6$\hbar\Omega$ when using 
the three-body effective interaction.

\subsection{Can Modern Nuclear Hamiltonians Tolerate a Bound Tetraneutron?}

\begin{figure}[b!]
\centering
\includegraphics[height=5.00in,angle=270]{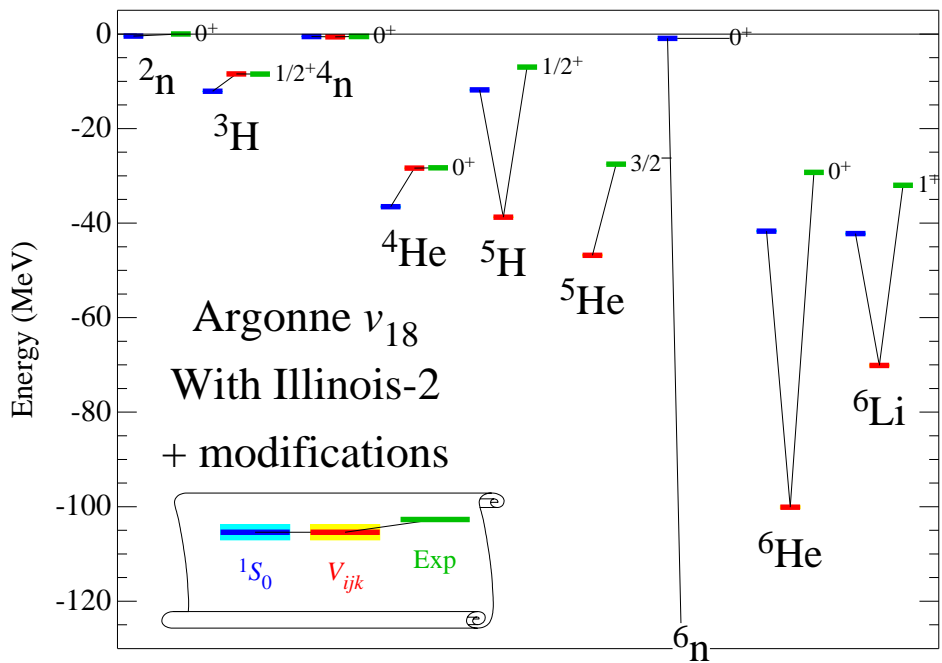}
\caption{Predictions of Hamiltonians that have been modified to bind
a tetraneutron.}
\label{fig:4n}
\end{figure}

An experimental claim of the existence of a bound tetraneutron cluster
($^4$n) was made recently~\cite{4n-exp}.  This prompted our
study~\cite{P03} of the theoretical possibility of such a state; the
conclusion was that a bound tetraneutron is strongly excluded by modern
nuclear Hamiltonians.  As a first step, negative-energy $^4$n solutions
using the AV18+IL2 model were searched for; GFMC calculations, using
propagation to very large imaginary time ($\tau=1.6$~MeV$^{-1}$),
produced only positive energies that steadily decreased as the RMS
radius of the system increased.  Adding artificial external wells to the
Hamiltonian can, of course, produce negative energies.  By varying the
well depth, one can extrapolate to zero-well depth.  Such calculations
suggest that there might be a $^4$n resonance near +2~MeV, but since the
GFMC calculation with no external well shows no indication of
stabilizing at that energy, the resonance, if it exists at all, must be
very broad.

To study the possibility that a minor modification of the AV18+IL2 model
could result in binding of $^4$n, a number of modifications to the AV18+IL2
model were made.  In each case the modification was adjusted to bind
$^4$n with an energy of approximately $-0.5$~MeV; the consequences of
this modification for other nuclei were then computed.
Figure~\ref{fig:4n} shows some of these results.  In the first such
modification, the AV1$^\prime$ potential~\cite{WP02} was used in just
the $^1\!S_0$ partial wave (and AV18 in the other partial waves); this
results in a $^4$n energy of $-0.52$~MeV.  However, as is shown by the
bars labeled $^1\!S_0$ in the figure, this results in substantial
overbinding of $^3$H, $^4$He and all other nuclei that were computed.
In addition it results in a bound dineutron; in fact the $^4$n is not
really bound as it can decay into two dineutrons.

\begin{figure}[b!]
\begin{minipage}[t]{78mm}
\includegraphics[height=78mm,angle=270]{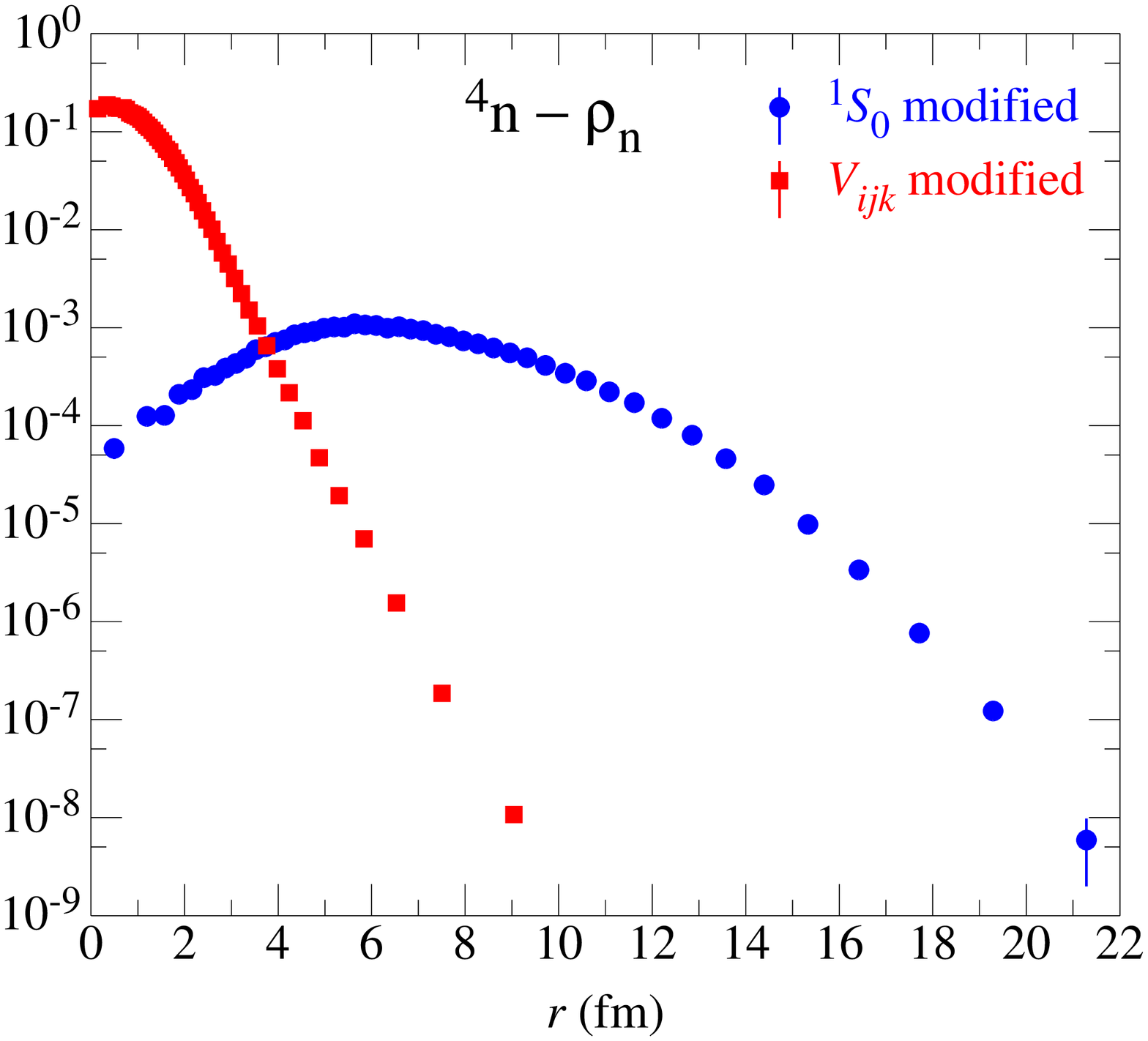}
\end{minipage}
\hspace{\fill}
\begin{minipage}[t]{80mm}
\includegraphics[height=80mm,angle=270]{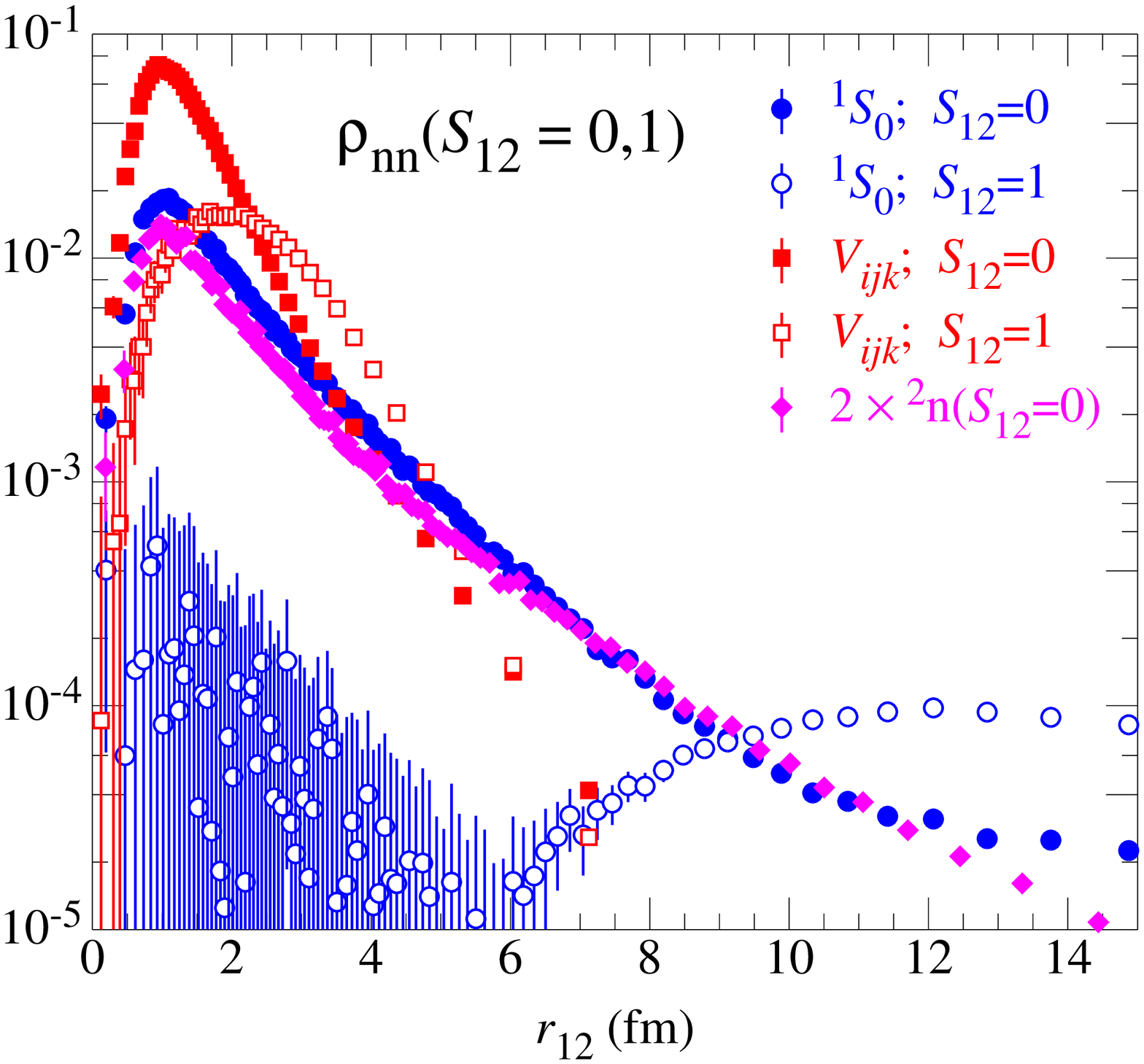}
\end{minipage}
\caption{One- and two-body densities of $^4$n bound with either a modified 
$^1\!S_0$ (circles) or $T=3/2$ $N\!N\!N$ potential (squares).
The diamonds show twice the two-body density of $^2$n.  Two-body
densities are shown projected into total spin 0 or 1 channels.}
\label{fig:4n_rho}
\end{figure}

Modifications to the $N\!N\!N$ or $N\!N\!N\!N$ potentials, which are
experimentally much less constrained than the $N\!N$ potential, could be
used to bind $^4$n. A $N\!N\!N$ potential that acts only in
$T=\case{3}{2}$ triples would have the same effect on $^4$n as one with
no isospin dependence, but would have no effect on $^3$H and $^4$He because they
contain only $T=\case{1}{2}$ triples.  Such a potential was added to the
AV18+IL2 Hamiltonian, and the coupling constant chosen to produce $^4$n
with $\sim -0.5$~MeV energy. It turns out that the coupling must be
quite large to produce the minimally bound $^4$n.  

This can be understood as follows.  If a modified $N\!N$ potential is
used to bind $^4$n, the pairs can sequentially come close enough to feel
the attraction; this allows the four neutrons to be in a diffuse, large
radius, distribution.  However a $N\!N\!N$ potential requires three
neutrons to simultaneously be relatively close and thus the density of
the system must be much higher.  Indeed the RMS radii of the $^4$n for
the modified $N\!N\!N$ potential is only 1.9~fm while the RMS
radius for $^4n$ produced by the modified $N\!N$ potential is 10~fm.
The left panel of Fig.~\ref{fig:4n_rho} shows the corresponding
densities.  The right panel compares the pair distributions for the two
different $^4$n systems with that of the dineutron made by the modified
$N\!N$ potential.  The $S=0$ pairs in the $^4$n made by the modified
$N\!N$ potential are basically undisturbed $^2$n pairs, while the
modified $N\!N\!N$ potential results in very different $^2$n pairs.

The small radius of the $^4$n bound by the modified  $N\!N\!N$ potential
results in a kinetic energy that is an order of magnitude more than for
the $^4$n bound by the modified $^1\!S_0$ potential.  This large kinetic
energy must be overcome by a large $N\!N\!N$ potential energy, and hence
a large coupling constant is required.

The very large coupling constant for the extra $N\!N\!N$ term
means that it has a large, even catastrophic, effect on any nuclear
system in which it can act.  This is shown in
Fig.~\ref{fig:4n} by the bars labeled $V_{ijk}$; for example
the binding energy of
$^6$Li is doubled and that of $^6$He is  tripled.
As stated above, this potential has no effect on $^3$H or $^4$He.  
However the most dramatic result of this potential is that every
investigated pure neutron system with $A>4$ is extremely bound and in
fact is the most stable ``nucleus'' of that A; the $^6$n energy
is off the scale of the figure at -650~MeV!

A four-nucleon potential that acts in only $T=2$ quadruples was
also constructed to weakly bind $^4$; it has even more devastating
consequences for nuclei in which it can act than the $N\!N\!N$ potential.
Thus an experimental observation of a bound tetraneutron would not
be a minor perturbation to our understanding of nuclei; the experimental
evidence for such a claim should be very strong before it is taken
seriously.

\subsection{RMS radii of helium isotopes}

\begin{figure}[t!] 
\hspace*{\fill}\includegraphics[scale=.585,angle=270]{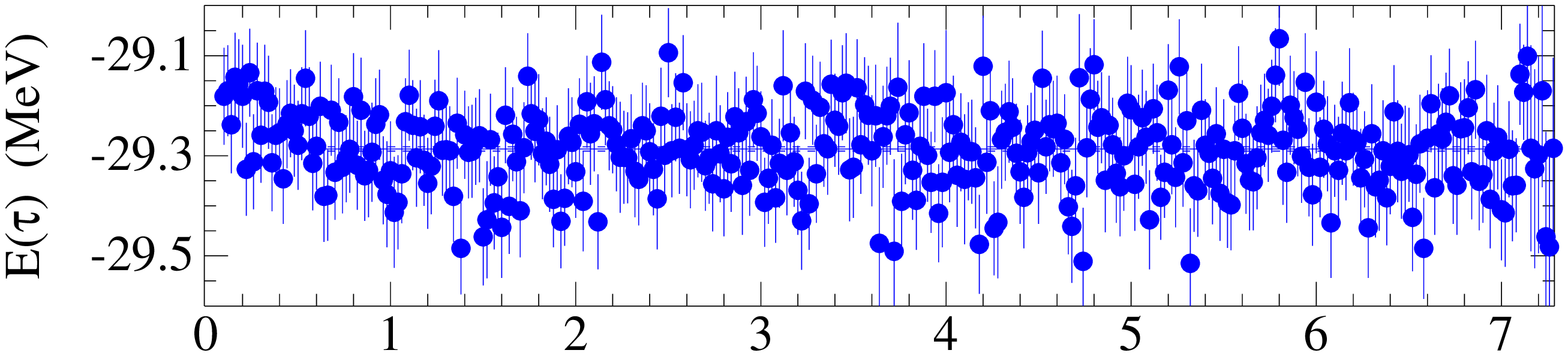}
\hspace*{\fill}\includegraphics[scale=.585,angle=270]{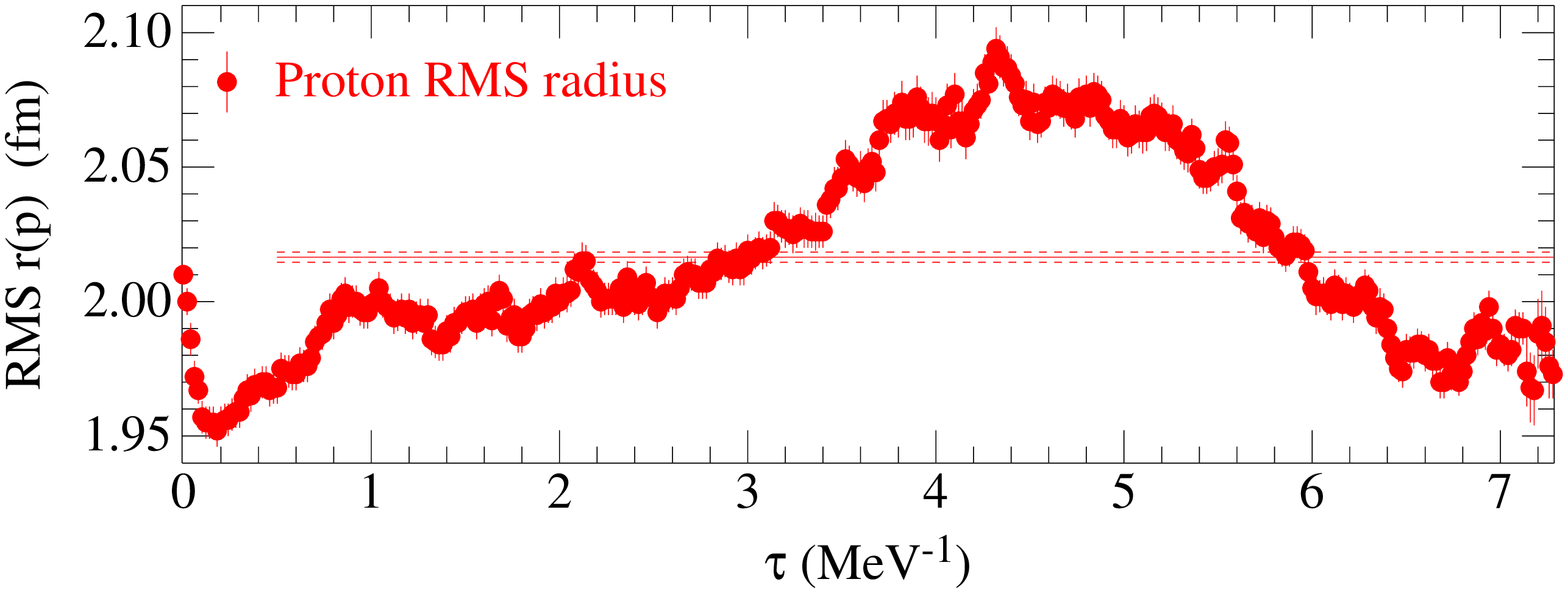}
\caption{GFMC propagation of $^6$He energy (top panel) and RMS
point proton radius (bottom panel).}
\label{fig:rmsr}
\end{figure}

Recently a group at Argonne has measured the RMS charge radius of the
radioactive nucleus $^6$He ($\beta$-decay halflife 0.8 sec.) with the
remarkable accuracy of 0.7\%~\cite{6he-rms}.  This prompted us to revisit
our already published~\cite{PPWC01} proton RMS radius for $^6$He.  That
calculation had been made with our standard propagation to
$\tau=0.2$~MeV$^{-1}$ and had produced a value in excellent agreement
with the new experimental value.  We made new calculations to much
larger $\tau$ and found that there are very slow fluctuations of the
RMS radius with $\tau$.  These are shown in Fig.~\ref{fig:rmsr} which
contains a GFMC propagation to $\tau=7.25$MeV$^{-1}$ for $^6$He using
the AV18+IL2 Hamiltonian.  The upper panel shows the energy on a highly
magnified scale; the results show just Monte Carlo statistical
fluctuations with no long-term correlations.  The standard error of the
mean that is extracted from these numbers is thus reliable.  However the
RMS point proton radius shown in the lower panel has extremely
long-range correlations making the error of the mean a useless number.
We do not understand the origin of these fluctuations and have been
unable to reduce them by using different importance sampling methods in
the GFMC propagation.  At the moment all we can do is report the average
shown in the figure and use the extrema of the fluctuations as an
estimate of the error; after folding in the proton and neutron RMS radii,
this is reported in Table~\ref{table:rms} as a
charge radius.  As can be seen our computed value is significantly above
the experimental value.  We have also made a large-$\tau$ calculations
of the RMS radii of $^4$He and $^8$He.  The $^4$He calculation shows no
long-term fluctuations and that of $^8$He shows substantially smaller
fluctuations than for $^6$He.  These results are also reported in the
table.

\begin{table}[ht!]
\caption{Computed and experimental RMS charge radii (in fm) of helium isotopes}
\label{table:rms}
\renewcommand{\arraystretch}{1.2} 
\renewcommand{\tabcolsep}{2pc} 
\begin{tabular}{lll}
\hline
         &  AV18+IL2  &  ~~Expt.  \\
\hline
$^4$He   & ~1.660(10) &  1.673(1) \\
$^6$He   & ~2.15(7)   &  2.054(14) \\
$^8$He   & ~1.98(4)   &        \\
\hline
\end{tabular}
\end{table}

\subsection{Center of Mass Effects on Spectroscopic Factors}

Spectroscopic factors are a measure of probability of finding
subcluster structure in a nucleus.  For the case of a specific
$A$-1 nuclear state in an $A$-body nucleus they are defined by
computing the quasi-hole wave function for nucleon removal:
\begin{equation}
\chi(r) = \langle [\Psi_{A-1}(J^\prime) \times N(\ell_j)]_{_J} | a(r)
| \Psi_{A}(J) \rangle~~~.
\end{equation}
The spectroscopic factor is then
\begin{equation}
\mathcal{S} = \int dr r^2 |\chi(r)|^2~~~.
\end{equation}
It is straightforward to compute $\mathcal{S}$ from shell-model
wave functions.  
Conventional shell model calculations are done in a fixed center (FC) 
with harmonic oscillator wave functions.  
Dieperink and de Forest~\cite{ddf} showed how to convert such
$\mathcal{S}$ to the desired translationally invariant (TI) ones.
For the case of removing a single nucleon from the $p$ shell,
their formula becomes
\begin{equation}
\mathcal{S}_\mathtt{TI} = \frac{A}{A-1}\mathcal{S}_\mathtt{FC}~~~.
\label{fcti}
\end{equation}
This increase of $\mathcal{S}$ arises from the center of mass
acquiring a $L=1$ component from the $p$-shell nucleons.  This
means that the nominally $s$-shell core nucleons have some
$p$-wave component and thus contribute to the $p$-wave spectroscopic
factors; $s$-wave spectroscopic factors are reduced to conserve
the total number of nucleons.

This theorem assumes that the same oscillator parameter 
is being used for $s$- and $p$-shell nucleons.
However, this is not a realistic assumption; if the oscillator 
parameter is chosen to give a good RMS radius for the $s$-shell
core ($^4$He) then very small RMS radii will be obtained
for the $p$-shell nuclei.  An example is shown in Table~\ref{table:specfac}
which shows several calculations of spectroscopic factors for
removing a proton from the ground state of $^7$Li.  
The rows show various cases from an uncorrelated shell model to
experimental values.  The columns show the point RMS radii of $^7$Li
and the spectroscopic factors computed in a fixed center and with
translationally-invariant wave functions.  For each case the
spectroscopic factors to the ground state (0$^+$) of $^6$He and the
sum of the factors for the 0$^+$ and 2$^+$ states are shown.
The final column shows the ratio of the TI to FC sums.

\begin{table}[tb!]
\caption{Spectroscopic factors for $^7$Li($\case{3}{2}^-$) $\rightarrow$
$^6$He + $p$: $0^+ + p_{\case{3}{2}}$, $~~2^+ + p_{\case{1}{2}}$, and $~~2^+
+ p_{\case{3}{2}}$ computed in a fixed center (FC) and with
translational invariance (TI).}
\label{table:specfac}
\renewcommand{\arraystretch}{1.2} 
\renewcommand{\tabcolsep}{2pc} 
\begin{tabular}{ccccc}
\hline
$a_{\mathtt{osc}}$ & $^7$Li $\langle r^2 \rangle^{\case{1}{2}}$ &
                               $\mathcal{S}_\mathtt{FC}$ &
                                            $\mathcal{S}_\mathtt{TI}$ &
                                                          $\mathtt{TI}/\mathtt{FC}$   \\

  $S~~~~P$      &    $p~~~~~n$ & 0$^+$ ~$\Sigma$  & 0$^+$ ~$\Sigma$  &  \\
\hline
 1.95 1.95  &  1.76 ~1.84 &  .56 ~ 1.~  &  .65 ~1.17  & \case{7}{6}   \\
 1.95 2.5~  &  1.99 ~2.16 &  .56 ~ 1.~  &  .63 ~1.13  &    1.13  \\
 1.95 3.0~  &  2.22 ~2.47 &  .56 ~ 1.~  &  .59 ~1.04  &    1.04  \\
 Correlated &  2.27 ~2.48 &  .31 ~ 0.56 &  .33 ~0.60  &    1.07  \\
 Correlated &  2.41 ~2.50 &  .32 ~ 0.61 &  .30 ~0.54  &    0.89  \\
 VMC        &  2.33 ~2.48 &  ~$-$~~$-$~ &  .36 ~0.61  &    ~$-$~   \\
 Expt    & ~2.26 ``2.51'' &  ~$-$~~$-$~ &  .42 ~0.58  &    ~$-$~  \\
\hline
\end{tabular}
\end{table}

The first three rows are for one-body harmonic-oscillator wave functions
with no pair or triplet correlations; thus the FC results correspond to
standard shell-model wave functions.  The TI wave functions were made by
expressing the oscillator wave functions in terms of
$r_i-R_{\mathrm{CM}}$ where $r_i$ is a nucleon coordinate and
$R_{\mathrm{CM}}$ is the center of mass of all $A$ or $A$-1 nucleons.
The oscillator parameter for the $s$-shell gives a good RMS radius for
$^4$He.  This same parameter is used for the $p$-shell in the first line
and results in very small $^7$Li RMS radii.  In this case the full
$A/(A-1)$ ratio of TI to FC spectroscopic factors is observed, as is
required by Eq.~(\ref{fcti}).  In the next two lines, the $p$-shell
oscillator parameter is increased until reasonable $^7$Li radii are
achieved; this results in significantly reduced TI/FC ratios.  The two
lines labeled ``Correlated'' report results from modified VMC wave
functions which can be expressed in either the FC or TI.  
These have the structure of Eqn.~(\ref{eq:psit}-\ref{Phi_A}) but
also include one-body wave functions for the $\alpha$-core nucleons.
Here we see
TI/FC ratios near 1 or even less than 1.  The line labeled ``VMC'' shows
the TI results for our full variational wave function; this wave
function has no useful FC form.  Based on these results it seems better
to make no conversion of a FC spectroscopic factor computed with just a
$p$-shell shell model, rather than increase it by $A/(A-1)$.  We are
continuing to study this.

\begin{figure}[b!]
\centering
\includegraphics[height=5.00in,angle=270]{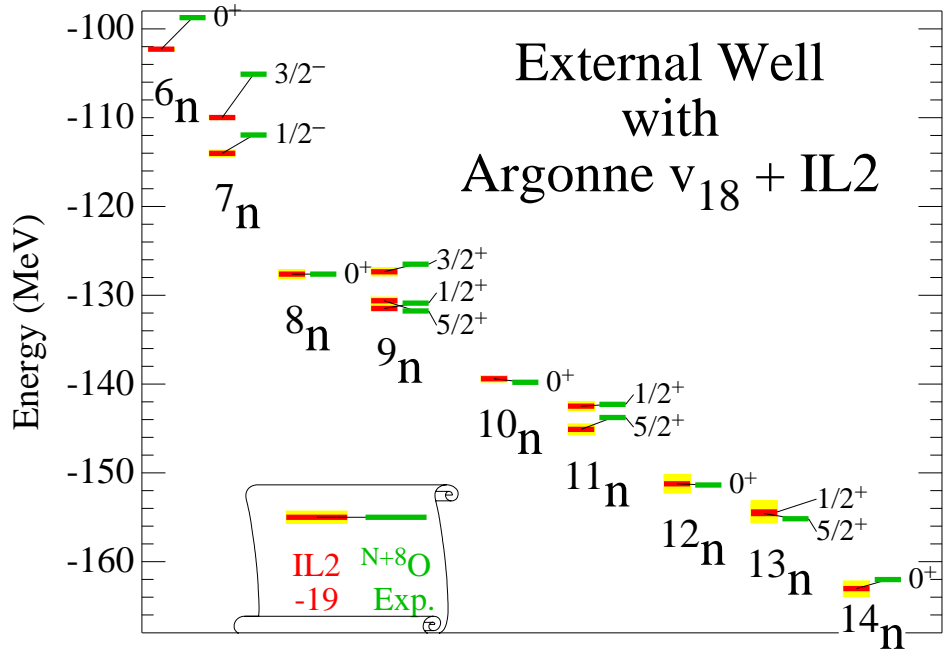}
\caption{Energies of neutron drops compared to the corresponding oxygen isotope
experimental energies.}
\label{fig:ndrops}
\end{figure}

\subsection{Mimicking Oxygen Isotopes as Neutron Drops}

Although GFMC is limited to light nuclei, it can be used to indirectly
study large, neutron-rich, nuclei by computing neutron drops.  These are
collections of neutrons interacting via a realistic Hamiltonian, such as
AV18+IL2, with the addition of an artificial external one-body well.  
The well provides the additional attraction necessary to produce
a bound state of the $N$ neutrons; it can be thought of as the
average effect of the protons which are not explicitly included
in the calculation.  Note that the $N$ neutrons occupy all
shells from 0$s$ up; the well does not represent any omitted
neutrons.

Figure~\ref{fig:ndrops} shows results for one such study in which
we attempt to mimic the neutron-rich isotopes of oxygen.  
Here we consider that the $N$-neutron drop represents $^{8+N}$O.
The 
external well is of the Woods-Saxon form with parameters
$R=3.0$ fm, $a=1.1$ fm, and $V=-35.5$ MeV; these were chosen
in an attempt to get the correct separation energies for $^{17,18}$O.
Since the energy of the 8 protons is not included in the calculation,
the energies of all the neutron drops are shifted (by 19 MeV) to 
match the 8-neutron drop with the experimental $^{16}$O energy.
The Woods-Saxon well binds 0$s$ and 0$p$ nucleons but does not
have 0$d$ or 1$s$ bound states; thus the binding of neutrons
in the $sd$ shell arises from the combination of the well and
the intrinsic AV18+IL2 potentials.

We see that the energies of the neutron-rich oxygen isotopes up to
$^{22}$O are reasonably well reproduced by this model.  However the
9-neutron drop has a $\case{1}{2}^+$ ground state rather than the
desired $\case{5}{2}^+$.  It appears necessary to add a spin-orbit term
to the well to achieve the correct ground state.  Once a satisfactory
model has been produced, the densities and other properties of these
model oxygen isotopes can be studied.

\section{Conclusions}

  Calculations with errors of only 1 to 2\% of energies for nuclei from 
$A$ = 6 to 12 for a given Hamiltonian are now possible.  The AV18+IL2
Hamiltonian gives average binding-energy errors $< 0.7$ MeV for $A = 3-12$.
A three-nucleon potential is required to obtain sufficient binding in the $p$-shell;
it is also required to reproduce many experimental spin-orbit splittings 
and several level orderings.  The GFMC and VMC wave functions can be
used to study many other nuclear properties.


\begin{thebibliography}{9}
\bibitem{PW01}
S.~C.~Pieper and R.~B.~Wiringa,
Annu. Rev. Nucl. Part. Sci. \textbf{51} (2001) 53.

\bibitem{PVW02}
S.~C. Pieper, K. Varga, and R.~B. Wiringa
Phys. Rev. C \textbf{66} (2002) 044310.

\bibitem{PPWC01}
S.~C. Pieper, V.~R. Pandharipande, R.~B. Wiringa, and J. Carlson,
Phys. Rev. C \textbf{64} (2001) 014001.

\bibitem{PWC04}
S.~C. Pieper, R.~B. Wiringa, and J.~Carlson, Phys. Rev. C, in press.

\bibitem{nocore}
P. Navr\'atil and W. E. Ormand, Phys. Rev. C \textbf{68} (2003) 034305.

\bibitem{P03}
S.~C. Pieper, Phys. Rev. Lett. \textbf{90} (2003) 252501.

\bibitem{WSS95}
R. B. Wiringa,  V. G. J. Stoks, and  R. Schiavilla,
Phys. Rev. C \textbf{51} (1995) 38.

\bibitem{Nijmegen}
J. R. Bergervoet, et al. 
Phys. Rev. C \textbf{41} (1990) 1435;
V. G. J. Stoks., et al.
Phys. Rev. C \textbf{48} (1993) 792.

\bibitem{PPCW95}
B. S. Pudliner, V. R. Pandharipande, J. Carlson, and R. B. Wiringa,
Phys. Rev. Lett. \textbf{74} (1995) 4396.

\bibitem{FM57}
J. Fujita and H Miyazawa,
Prog. Theor. Phys. \textbf{17} (1957) 360.

\bibitem{PPCPW97}
B. S. Pudliner et al.,
Phys. Rev. C  \textbf{56} (1997) 1720.

\bibitem{Wir00}
R. B. Wiringa, S. C. Pieper, J. Carlson, and V. R. Pandharipande,
Phys. Rev. C \textbf{62} (2000) 014001.

\bibitem{KNG+01}
H. Kamada, et al.
Phys. Rev. C \textbf{64} (2001) 044001.

\bibitem{PP93}
S. C. Pieper and V. R. Pandharipande,
Phys. Rev. Lett. \textbf{70} (1993) 2541.

\bibitem{kurath56}
D. Kurath, Phys. Rev. \textbf{101} (1956) 216.

\bibitem{tmp}
S. A. Coon and H. K. Han, Few-Body Syst. \textbf{30} (2001) 131.

\bibitem{4n-exp}
F. M. Marqu\'es \textit{et al.}, Phys. Rev. C \textbf{65} (2002) 044006.


\bibitem{WP02}
R.~B. Wiringa and S.~C. Pieper,
Phys. Rev. Lett. \textbf{89},  (2002)182501.

\bibitem{6he-rms}
L.-B. Wang et al., Phys. Rev. Lett. \textbf{93} (2004) 142501.

\bibitem{ddf}
A. E. L. Dieperink and T. de Forest, Jr., Phys. Rev. C \textbf{10} (1974) 543.

\end{thebibliography}
\end{document}